\documentclass[preprint,twocolumn,]{aastex63}
\hypersetup{linkcolor=red,citecolor=green,filecolor=cyan,urlcolor=magenta}
\usepackage{amsmath}

\received{xx xxx}
\revised{xx xxx}
\accepted{xx xxx}
\submitjournal{ApJ}

\shorttitle{Quarks, Failing CCSN}
\shortauthors{Zha et al.}

\graphicspath{{./}{figures/}}

\begin{document}
	
	\title{Progenitor Dependence of Hadron-quark Phase Transition in Failing Core-collapse Supernovae}
	
	\correspondingauthor{Shuai Zha}
	\email{shuai.zha@astro.su.se}
	
	\author[0000-0001-6773-7830]{Shuai Zha}
	\affiliation{The Oskar Klein Centre, Department of Astronomy,\\ Stockholm University, AlbaNova, SE-106 91 Stockholm, Sweden}

	\author[0000-0002-8228-796X]{Evan P. O'Connor}
	\affiliation{The Oskar Klein Centre, Department of Astronomy,\\ Stockholm University, AlbaNova, SE-106 91 Stockholm, Sweden}
	
	\author[0000-0003-0849-7691]{André da Silva Schneider}
	\affiliation{The Oskar Klein Centre, Department of Astronomy,\\ Stockholm University, AlbaNova, SE-106 91 Stockholm, Sweden}
	
	\begin{abstract}
		We study the consequences of a hadron-quark phase transition (PT) in failing core-collapse supernovae (CCSNe) which give birth to stellar-mass black holes (BH). We perform a suite of neutrino-transport general-relativistic hydrodynamic simulations in spherical symmetry with 21 progenitor models and a hybrid equation of state (EoS) including hadrons and quarks. We find that the effect of the PT on the CCSN postbounce dynamics is a function of the bounce compactness parameter $\xi_{2.2}$. For $\xi_{2.2}\gtrsim0.24$, the PT leads to a second dynamical collapse of the protocompact star (PCS). While BH formation starts immediately after this second collapse for models with $\xi_{2.2}\gtrsim0.51$, the PCS experiences a second bounce and oscillations for models with $0.24\lesssim\xi_{2.2}\lesssim0.51$. These models emit potent oscillatory neutrino signals with a period of $\sim$ms for tens of ms after the second bounce, which can be a strong indicator of the PT in failing CCSNe if detected in the future. However, no shock revival occurs and BH formation inevitably takes place in our spherically-symmetric simulations. Furthermore, via a diagram of mass-specific entropy evolution of the PCS, the progenitor dependence can be understood through the appearance of third-family of compact stars emerging at large entropy induced by the PT.
	\end{abstract}
	
	
	\section{Introduction}
	The death of a massive star can give birth to a stellar-mass black hole (BH) if it fails to explode as a core-collapse supernova (CCSN) \citep{2006PhRvL..97i1101S,2011ApJ...730...70O}. The process of this BH formation channel has been studied with neutrino-transport hydrodynamic simulations in great detail \citep{2007ApJ...667..382S,2011ApJ...730...70O,2018MNRAS.477L..80K,2020arXiv201002453P,2020PhRvD.101l3013W,2020ApJ...894....4S}. As in a successful CCSN, the collapse of the progenitor's iron core produces a protocompact star (PCS) which rebounds due to the stiffening of the nuclear matter equation of state (EoS) just above nuclear saturation density ($\rho_{\rm sat}\simeq2.7\times10^{14}$~g~cm$^{-3}$) \citep{2021Natur.589...29B}. However, in a failing CCSN, shock revival does not arise to unbind the stellar envelope. Thus, the PCS continuously grows through accretion and inevitably collapses into a BH once its mass exceeds the maximum compact star mass allowed by the EoS \citep{2020ApJ...894....4S}.
	
	The recent discovery of a $\sim2.6~M_\odot$ compact object via gravitational-wave (GW) detection \citep{2020ApJ...896L..44A} boost the quest to determine whether such massive compact stars contain a quark matter core \citep[e.g. ][]{2020PhRvL.125z1104T,2020arXiv201011020R,2020arXiv200900942C}. During the postbounce growth and contraction of a PCS, its central density can increase from $\sim\rho_{\rm sat}$ to $\gtrsim10\times\rho_{\rm sat}$. At such high densities (and also high temperatures), free quarks may become deconfined from hadrons, i.e. protons and neutrons \citep{1984PhRvD..30..272W}, in the PCS. Such a hadron-quark PT is found to induce collapse of the PCS to a small radius, and a second core bounce due to stiffening of the quark matter EoS may help to revive the supernova shock \citep{2009PhRvL.102h1101S,2018NatAs...2..980F,2020PhRvL.125e1102Z}. Previous focus on the PT has been devoted to the search of an appropriate EoS which simultaneously fulfills the constraint on the maximum mass of compact stars  \citep[$\gtrsim 2.0~M_\odot$, ][]{2010Natur.467.1081D,2013Sci...340..448A,2020NatAs...4...72C} and enables a successful CCSN explosion \citep{2016PhRvD..94j3008H,2017PhRvD..96e6024K}.
	
	On the other hand, it is also interesting to assess the consequences of a PT on the evolution toward failed supernovae. For a particular hybrid EoS, \cite{2013A&A...558A..50N} found that BH formation starts shortly after the PCS forms and collapse takes place without a second bounce: the PT has the effect of shortening the time between the first bounce and BH formation. However, because \cite{2013A&A...558A..50N} only simulated a single failing CCSN model, the full potential landscape is unclear.
	
	Currently there is no consensus on the locus of the PT in the QCD phase diagram \citep{2017RvMP...89a5007O,2020PhRvD.102g4017D,2020PhRvD.102l3023B}. Several phenomenological models \citep{1984PhRvD..30.2379F,1961PhRv..122..345N,2017PhRvD..96e6024K,2021PhRvD.103b3001B} have been used to calculate the quark matter EoS and the PT is constructed assuming either Maxwell or Gibbs conditions \citep{1992PhRvD..46.1274G}. An interesting property of the PT is the appearance of a secondary unstable branch in the mass-radius (M-R) curve of compact stars \citep{2013PhRvD..88h3013A}. Particularly, this may appear only at a large specific entropy for a class of hybrid EoSs including the PT \citep{2013PhRvD..88h3013A,2013AstL...39..161Y,2016PhRvD..94j3001H}. This property is believed to be important for the shock revival after the PT-induced collapse in CCSNe \citep{2016PhRvD..94j3001H}. From the CCSN side, it is known that the specific entropy of the PCS is related to the progenitor compactness and impacts the BH formation process \citep{2020ApJ...894....4S}. Therefore, the effect of the PT can be progenitor dependent and this has not been systematically studied before.
	
	Recently the hadron-quark PT has received more attention in nuclear astrophysics because future multi-messenger observations with GW \citep{2009MNRAS.392...52A,2019PhRvL.122f1102B,2019PhRvL.122f1101M,2020PhRvL.125e1102Z,2020PhRvD.102l3023B} and neutrinos \citep{2009PhRvL.102h1101S,2010PhRvD..81j3005D,2010ApJ...721.1284N,2018NatAs...2..980F} can provide invaluable information on this aspect. Bursts of GWs \citep{2020PhRvL.125e1102Z} and neutrinos \citep{2009PhRvL.102h1101S} are predicted in computational simulations related to the PT-induced collapse and bounce of the PCS inside successful CCSNe. A failing CCSN may shine less in electromagnetic waves but comparable in GWs and neutrinos compared to a successful one \citep{2006PhRvL..97i1101S,2011PhRvL.106p1103O,2018ApJ...857...13P,2020PhRvD.101l3013W}. Whether these signals carry information on the occurrence of a hadron-quark PT in a failing CCSN has not yet been explored.
	
	In this paper, we investigate the consequences of a hadron-quark PT in failing CCSNe using a suite of spherically-symmetric simulations with 21 progenitor models. The paper is organized as follows. In Section~\ref{sec:meth} we described the EoS and simulation setup, including the progenitor models and computational code. We present the results of our simulations, including the postbounce dynamics and neutrino signals, in Section~\ref{sec:res}. We also interpret the progenitor dependence based on the properties of EoS in this section. We conclude in Section~\ref{sec:con}.
	
	\section{Methods \label{sec:meth}}
	\subsection{Equation of state}
	To study the hadron-quark PT, we use a hybrid EoS \citep{2010JPhG...37i4064S} composed of the STOS EoS \citep{1998PThPh.100.1013S} for hadronic matter and the MIT bag model EoS \citep{1984PhRvD..30.2379F} for quark matter. The PT region is constructed under the Gibbs conditions which allow different charge fractions for the hadronic portion and quark portion in the mixed phase \citep{1992PhRvD..46.1274G}. This leads to monotonically increasing pressure in the PT region as the fraction of quarks (and matter density) increases. Note that there is currently no consensus on the treatment of charge neutrality in the mixed phase \citep{2017RvMP...89a5007O} and such a construction should be treated as a limiting case.
	
	The bag constant $B$ of the hybrid EoS is chosen to be 145~MeV, which is the lower limit given by hadron fitting \citep{1983ARNPS..33..235D}. Additionally, a strong coupling constant $\alpha_s=0.7$ is included as in Eq.~(9) of \cite{2011ApJS..194...39F}. As a result, the maximum gravitational mass of cold and $\beta$-equilibrium compact stars is $\sim2.0~M_\odot$ for the hybrid EoS, which is marginally consistent with the stringent constraint from the pulsar mass in \cite{2013Sci...340..448A}. The PT region is from $2.5\times10^{15}$~g~cm$^{-3}$ to $5.0\times10^{15}$~g~cm$^{-3}$ for cold and symmetric (electron fraction $Y_e=0.5$) matter, and from $4.0\times10^{14}$~g~cm$^{-3}$ to $4.0\times10^{15}$~g~cm$^{-3}$ at $T=10~{\rm MeV}$ and $Y_e=0.3$ (a condition similar to that of the core of a PCS in CCSNe). 
	
\begin{figure}[t!]
		\centering
		\includegraphics[width=0.47\textwidth]{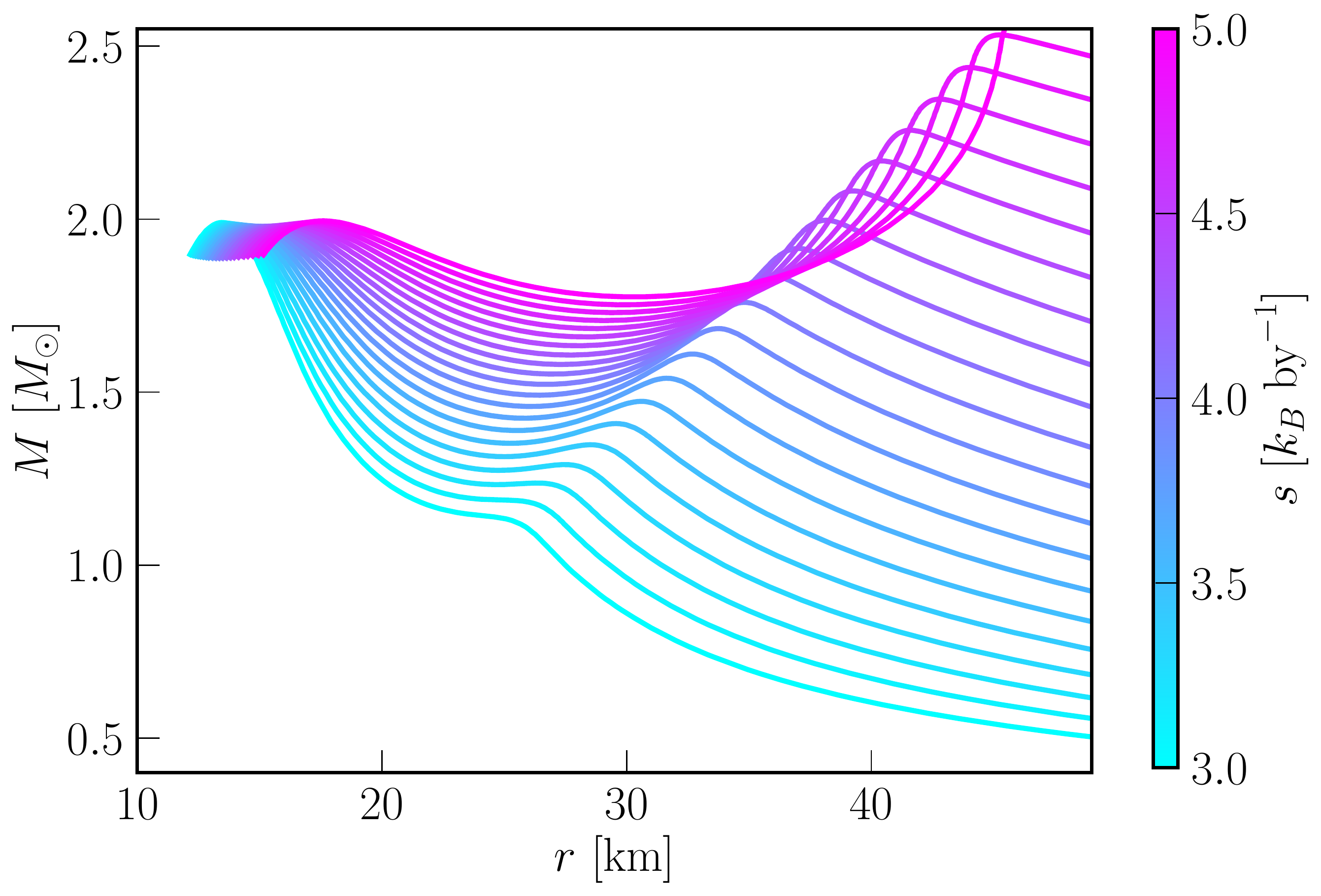}
		\caption{Mass-radius curves of compact stars with a constant specific entropy $s$ and in $\beta$-equilibrium. The line color is parametrized by entropy. \label{fig:mr}}
	\end{figure}

	 We are interested in the effect of the PT on CCSNe, whose PCSs are generally hot, with a specific entropy $s$ in the range of $2.5-5~k_B~{\rm baryon}^{-1}$ \citep{2020ApJ...894....4S}. In Fig.~\ref{fig:mr}, we plot the M-R curves of compact stars with different constant $s$ and in $\beta$-equilibrium for the hybrid EoS. For $s\gtrsim3.5~k_B~{\rm baryon}^{-1}$, the M-R curve has two extreme masses, connected by an unstable and a stable branch. This is the so-called ``both" third-family topology of \citep{2013PhRvD..88h3013A}. We denote the extreme mass with a smaller radius as $M_{3,\max}$ and the one with a larger radius as $M_{2,\max}$. $M_{3,\max}$ is almost independent of $s$ while $M_{2,\max}$ increases monotonically for increasing $s$. As a result, $M_{2,\max}$ becomes larger than $M_{3,\max}$ for $s\gtrsim 4.5~k_B~{\rm baryon}^{-1}$. This trend has been explained in detail by \citet{2016PhRvD..94j3001H} and is important to understand the effects of the PT in CCSNe.
	
	\subsection{Simulation setup}
	We use a suite of solar-metallicity progenitor models from \cite{2018ApJ...860...93S}. We choose 7 models evolved with a standard mass loss rate (dubbed sxx.xx) and 14 progenitor models with a mass loss rate reduced by a factor of 10 (dubbed sxx) \citep{2018ApJ...860...93S}. We expect our results to only depend on the iron-core structure of progenitor models but not on metallicity and mass loss rate. Therefore, the choice of model set is rather arbitrary and models are selected from different sets to span and explore the range of progenitor compactness.
	
	It has been pointed out that to first order, the trend of BH formation in failing CCSNe depends on the compactness parameter \citep{2011ApJ...730...70O}:
	\begin{equation}
	\xi_M=\frac{M/M_\odot}{R(M_{\rm baryon}=M)/1000~{\rm km}}\bigg|_{t=t_0}. \label{eq:xi}
	\end{equation}
	$\xi_M$ is a good indicator for the time between the core bounce and BH formation if $M$ is approximately the maximum baryonic mass of compact stars allowed by the EoS \citep{2011ApJ...730...70O,2020ApJ...894....4S}. We set $M=2.2~M_\odot$,corresponding to the maximum baryonic mass of cold compact stars possible from this hybrid EoS, and use the time at the iron-core bounce as $t_0$. All the progenitor models used in this work have $\xi_{2.2}\gtrsim0.1$ and collapse to BHs in a reasonably short time. Their mass and $\xi_{2.2}$ are listed in Table~\ref{tab:dyn}.
	
	We carry out spherically-symmetric simulations of CCSNe with the general-relativistic hydrodynamics code GR1D \citep{2010CQGra..27k4103O,2015ApJS..219...24O}, which includes an ``M1" scheme for the three-flavor neutrino transport. The resolution of the computational grid is uniform in the inner 20~km with a grid size of 300~m and becomes logarithmically increasing outside until several $10^4$~km. All the simulations include at least $2.5~M_\odot$ of baryonic material from the progenitors. We use 18 logarithmically-spaced energy groups to sample the neutrino distribution function. The lowest energy group centers at 1~MeV with a width of 2~MeV while the largest one centers at 280.5~MeV with a width of $\sim61$~MeV. The neutrino-matter interaction rates in the mixed and quark phases are calculated in the same way as those in the hadronic phase using NuLib \citep{2015ApJS..219...24O}. The nucleon chemical potentials are calculated from the chemical potentials of quarks by
	\begin{equation}
	\begin{aligned}
	\mu_n=\mu_u+2\mu_d, \\
	\mu_p = 2\mu_u+\mu_d, 
	\end{aligned}
	\end{equation} 
	where $\mu_n$ ($\mu_p$) is the chemical potential of neutron (proton), and $\mu_u$ ($\mu_d$) is the chemical potential of up (down) quark.
	At such high densities where quarks start to become unconfined, neutrinos must be trapped and in thermal and weak equilibrium with matter and, therefore, the detailed rates should not be important \citep{2011ApJS..194...39F}. We leave the improvement of the neutrino-quark interaction rates for a future study.
	
	\section{Results \label{sec:res}}
	
	\subsection{Postbounce dynamics \label{ssec:dyn}}
	
	\begin{table*}[t!]
	\caption{Results of our 21 CCSN simulations. Two sets of solar-metallicity progenitor models are used as initial conditions, with one using a standard mass loss rate (dubbed sxx.xx) and the other using a mass loss rate reduced by a factor of 10 (dubbed sxx). $M$ is the zero-age-main-sequence mass. $\xi_{2.2}|_{t=0}$ and $\xi_{2.2}|_{t=t_{1b}}$ are the compactness parameters at the onset of collapse and at the first bounce as defined in Eq.~\ref{eq:xi}. $t_{1b}$, $t_{2c}$ and $t_{\rm BH}$ are the times at the moments of the first bounce,  second collapse and BH formation, respectively. $M_{{\rm PCS}, 2c}$ and $M_{{\rm PCS, BH}}$ are the gravitational mass of the PCS at moments of the second collapse and BH formation, respectively. $M_{\rm PCS}$ is defined as the mass enclosed by the accretion shock. \label{tab:dyn}}
	\centering 
	\begin{tabular}{cccccccccc}
		\hline
		Model & $M~[M_\odot]$ & $\xi_{2.2}|_{t=0}$ &$\xi_{2.2}|_{t=t_{1b}}$ & $t_{1b}$~[s] & $t_{2c}-t_{1b}$~[s] & second bounce & $t_{\rm BH}-t_{2c}$~[s] & $M_{{\rm PCS},2c}~[M_\odot]$ & $M_{{\rm PCS, BH}}~[M_\odot]$ \\ \hline
		s15 & 15 & 0.125 & 0.125 & 0.202 & - & - & $>$2.5 & - & -\\
		s18 & 18 & 0.197 & 0.200 & 0.230 & 1.289 & No& $>$2.5 & 1.81 & - \\
        s18.44 & 18.44 & 0.222 & 0.224 & 0.209 & 1.340 & No  & $>$2.5 &  1.76 & - \\
		s19.81 & 19.81 & 0.250 & 0.251 & 0.222 & 0.947 & Yes & $>$2.5 &  1.83 & - \\
		s22 & 22 & 0.279 & 0.280 & 0.253 & 0.819 & Yes& 1.379 &  1.89 & 2.05\\
		s16 & 16 & 0.288 & 0.289 & 0.225 & 0.849 & Yes& 1.178 & 1.86 & 2.04  \\
		s23 & 23 & 0.341 & 0.348 & 0.256 & 0.747 & Yes& 0.815 & 1.92& 2.07 \\  
		s17 & 17 & 0.342 & 0.349 & 0.245 & 0.742 & Yes& 0.808 & 1.91 & 2.07\\
		s19 & 19 & 0.400 & 0.407 & 0.249 & 0.628 & Yes& 0.532 & 1.95 & 2.08 \\
		s24 & 24 & 0.403 & 0.410 & 0.267 & 0.626 & Yes& 0.541 & 1.97 & 2.09 \\
		s19.89 & 19.89 & 0.421 & 0.433 & 0.258 & 0.606 & Yes& 0.455 & 1.97 & 2.09 \\
		s22.39 & 22.39 & 0.426 & 0.442 & 0.269 & 0.585& Yes & 0.505 & 1.99 & 2.11\\
		s25 & 25 & 0.445 & 0.458 & 0.260 & 0.580 & Yes & $> 0.86^{a}$ & 1.99 & -  \\
		s22.31 & 22.31 & 0.462 & 0.479 & 0.269 & 0.553 & Yes & $> 0.78^{a}$ & 2.00 & - \\
		s22.38 & 22.38 & 0.481 & 0.499 & 0.274 & 0.533 & Yes & $> 0.75^{a}$ & 2.02 & - \\ 
		s22.21 & 22.21 & 0.497 & 0.522 & 0.277 & 0.509 & No & $8\times10^{-4}$ & 2.03 & 2.02 \\
		s21 & 21 & 0.529 & 0.561 & 0.280 & 0.470 & No& $6\times10^{-4}$ & 2.05 & 2.05 \\
		s26 & 26 & 0.536 & 0.569 & 0.268 & 0.472 & No& $6\times10^{-4}$ & 2.05 & 2.05 \\
		s20 & 20 & 0.567 & 0.609 & 0.287 & 0.447 & No& $7\times10^{-4}$ & 2.08 & 2.07 \\
		s30 & 30 & 0.681 & 0.781 & 0.309 & 0.356 & No& $5\times10^{-4}$ & 2.16 & 2.16 \\
		s33 & 33 & 0.703 & 0.789 & 0.299 & 0.343 & No& $5\times10^{-4}$ & 2.16 & 2.16 \\
		\hline 
	\end{tabular} 
    \leftline{$^{a}$ These models have not collapsed into a BH at the end of the simulations. See text in Section~\ref{ssec:neu} for the details.}
    \end{table*}
	
	\begin{figure}[t!]
		\centering \includegraphics[width=0.47\textwidth]{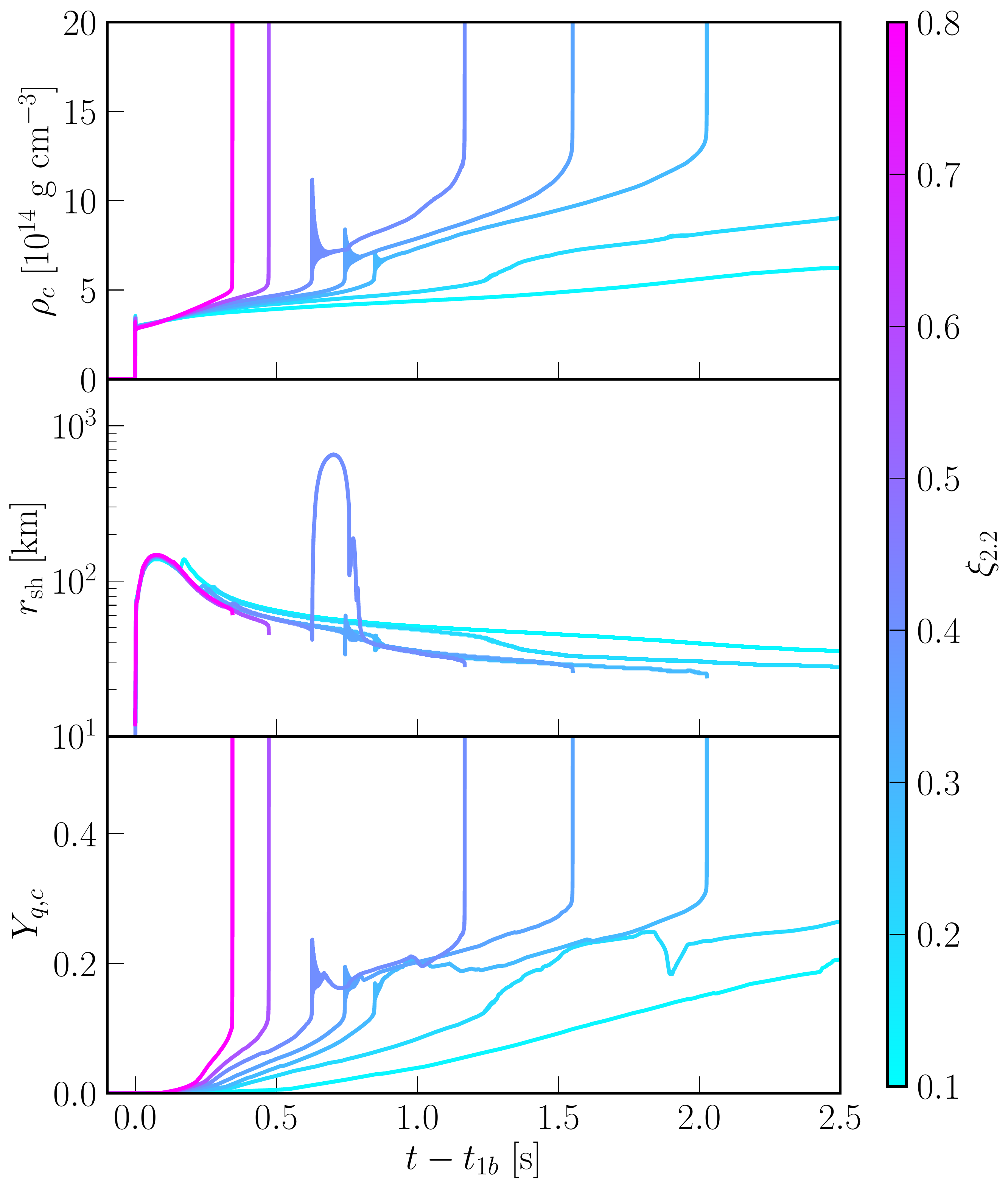}
		\caption{Time evolution of the central density (upper panel), shock radius (middle panel) and central quark fraction (lower panel) in 7 selected models. Color of the lines indicates the compactness parameter $\xi_{2.2}$ of each model \added{at the first bounce} and a cooler (hotter) color relates to a smaller (larger) $\xi_{2.2}$. From low to high compactness the models are s15, s18, s16, s17, s24, s26, s33. \label{fig:cen}}
	\end{figure}

	\begin{figure*}[t!]
		\centering
		\includegraphics[width=0.32\textwidth]{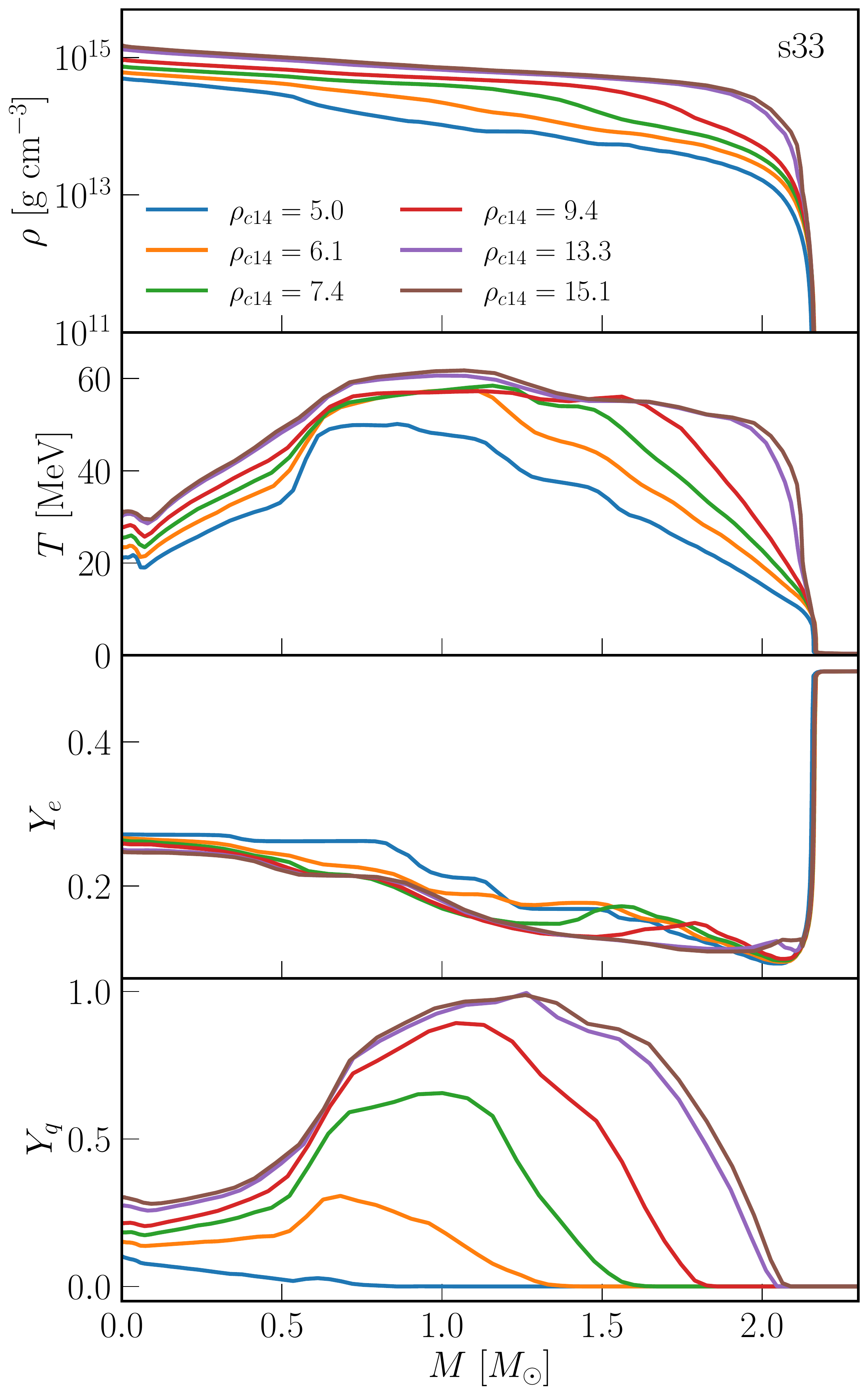}
		\includegraphics[width=0.32\textwidth]{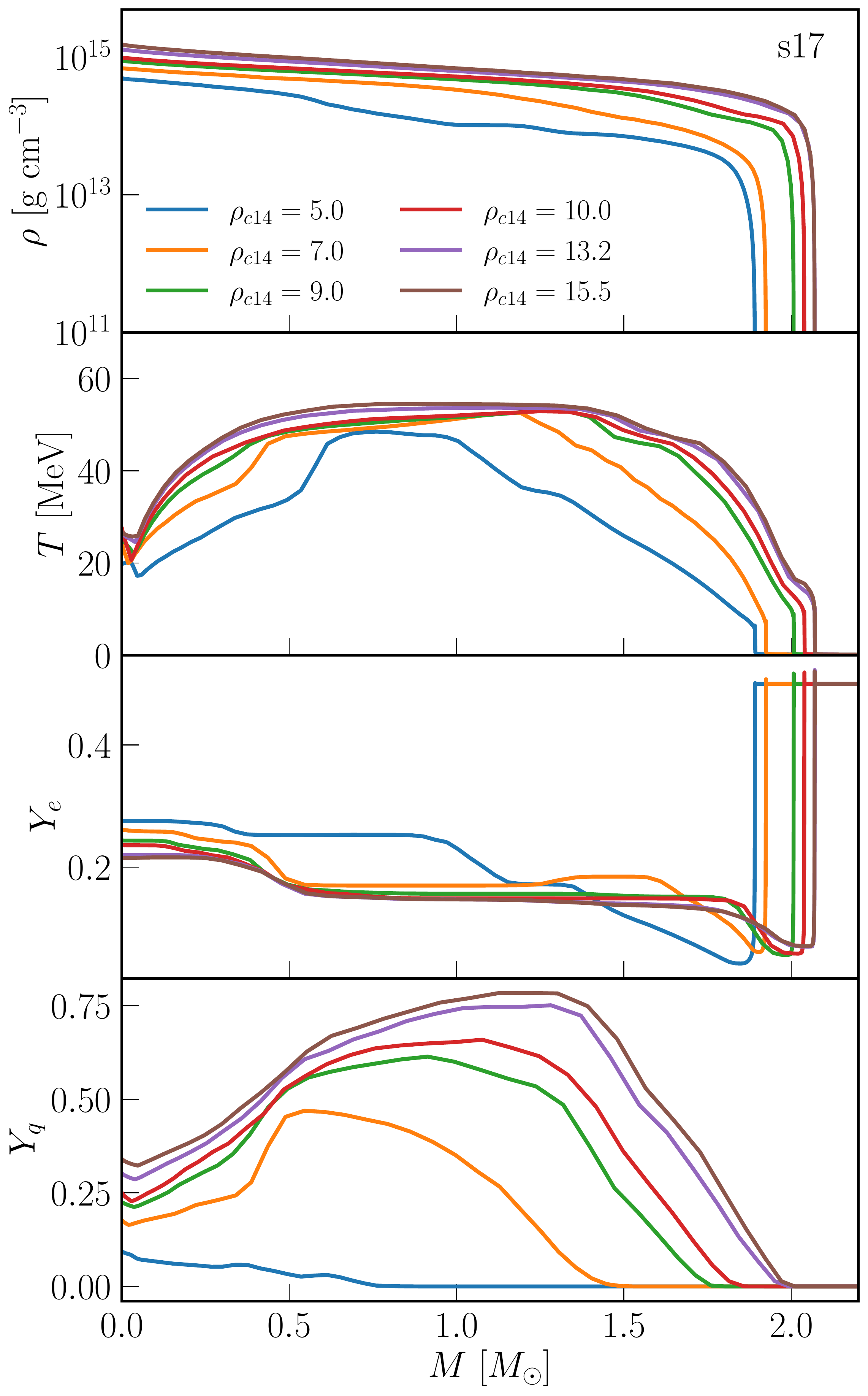}
		\includegraphics[width=0.328\textwidth]{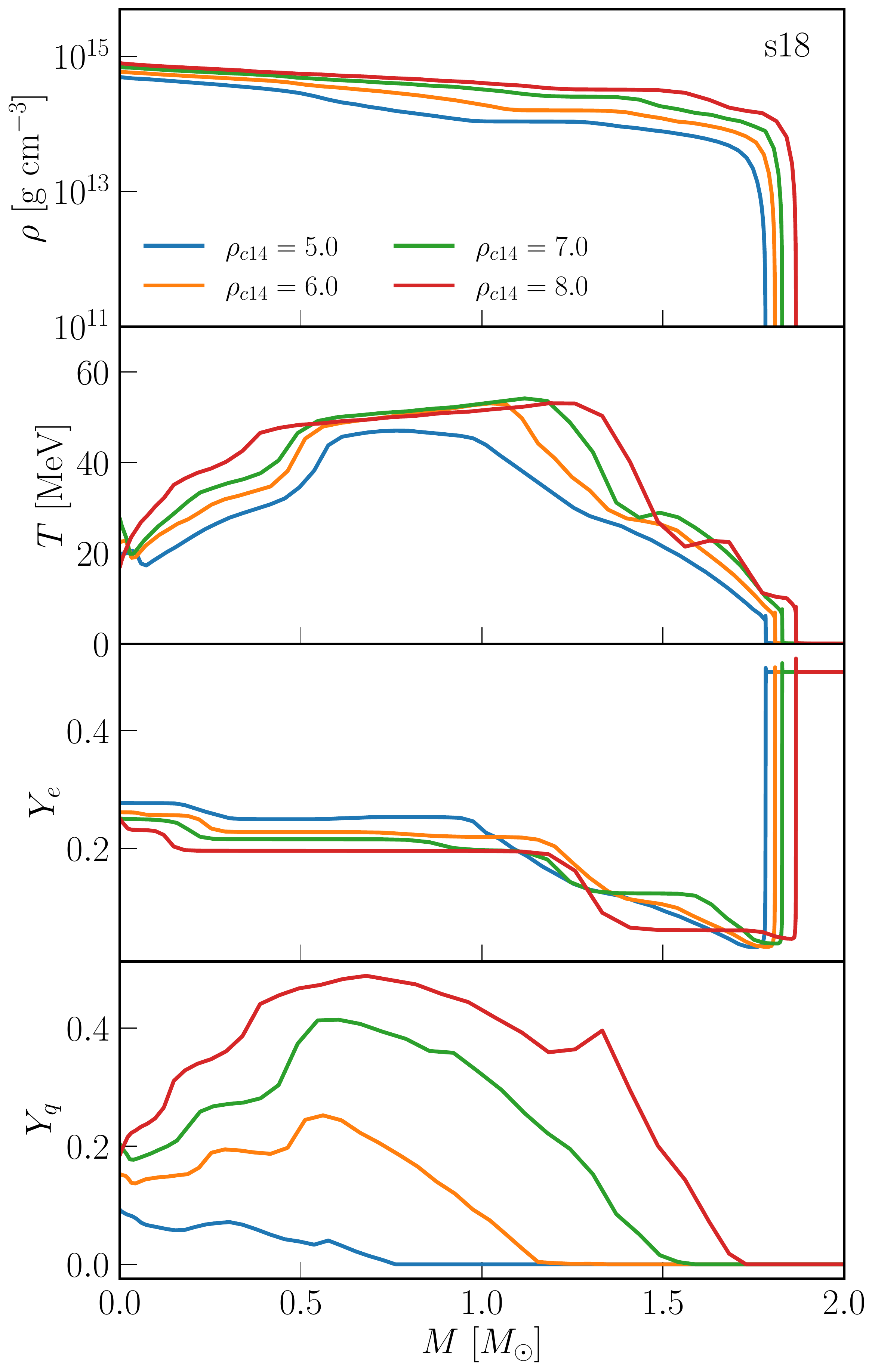}
		\caption{Density $\rho$, temperature $T$, electron fraction $Y_e$ and quark fraction $Y_q$ as a function of gravitational mass coordinate for models s33 (left), s17 (middle) and s18 (right). Color of the lines indicates the moment when the central density is $\rho_{c14}\times10^{14}~\rm g~cm^{-3}$. \label{fig:profiles} }
	\end{figure*}
	
	During the iron core collapse and early postbounce stages when the matter density is below $\rho_{\rm sat}$, the dynamics are not affected by the hadron-quark PT. For the progenitor models used in this work, central densities reach $\rho_{\rm sat}$ in $\sim0.2-0.3$~s after the onset of collapse. The core bounces due to stiffening of the hadronic matter EoS and we mark this moment as the first bounce at time $t_{1b}$. Then, the bounce shock quickly stalls and turns into an accretion shock. The PCS mass and central density increase continuously due to the accretion of matter though the shock.

	In Figure~\ref{fig:cen}, we plot the central density $\rho_c$, shock radius $r_{\rm sh}$ and central quark fraction $Y_{q,c}$ as a function of the postbounce time for 7 representative models. \added{The shock position is found by searching for the grid whose velocity differs from its neighboring grid by more than a certain value $\Delta v$ (0.01-0.02 times the speed of light). Because sometimes there are several shocks, we use the outermost one for $r_{\rm sh}$ to track the stalled bounce shock.} The onset of PT takes place $\sim0.1$~s after the first bounce signaled by the non-zero $Y_{q,c}$ at $\rho_c\simeq3.5\times10^{14}$~g~cm$^{-3}$ ($\sim1.3\rho_{\rm sat}$). Except for two models (s15 and s18) with the smallest $\xi_{2.2}$, $\rho_c$ experiences a second plunge phase in less than 1~s after the first bounce, accompanying with the receding of $r_{\rm sh}$. This indicates the collapse of the PCSs, which we mark as the second collapse. We use the moment when $\rho_c=6.0\times10^{14}$~g~cm$^{-3}$ with $Y_{q,c}\simeq0.12$ as the time of the second collapse $t_{2c}$.

	For the models with a large $\xi_{2.2}$ ($\gtrsim 0.51$), this second collapse directly leads to BH formation, as $\rho_c$ quickly increases and exceeds the maximum value of the EoS table. Similar to the results in \cite{2013A&A...558A..50N}, the hadron-quark PT shortens the time between the first bounce and BH formation. For example, the model s33 with the largest $\xi_{2.2}$, collapses to a BH $\sim$0.3 s ($\sim$0.8 s) after the first bounce with the hybrid (STOS) EoS. This is due to the softness and smaller maximum PCS mass of the hybrid EoS. The evolution of the PCS structure in the model s33 is shown in the left panel of Figure~\ref{fig:profiles}. The profiles are chosen when $\rho_c$ reaches a specific value $\rho_{c14}\times10^{14}~\rm g~cm^{-3}$. $Y_{q}$ has an off-center maximum after the onset of the second collapse which reaches $\sim1.0$ near BH formation. The off-center peak in $Y_{q}$ is related to the fact that a higher temperature and lower $Y_e$ lead to a lower density for the deconfinement of hadrons to quarks. We note that during the second collapse, $Y_e$ decreases quickly accompanying the increase of $Y_q$. This is due to the larger difference in chemical potentials $\mu_p-\mu_n$ ($\mu_u-\mu_d$) in the mixed phase, which leads to the depletion of electrons and the production of electron neutrinos. The gravitational mass of the PCS is $\sim2.16~M_\odot$ prior to the BH formation.
	
	For the less compact models ($\xi_{2.2}\lesssim 0.51$), the second collapse does not lead to BH formation immediately. Instead, $\rho_c$ reaches an extreme value, drops and then oscillates for tens of ms. In the models s16 and s17, $r_{\rm sh}$ oscillates together with $\rho_c$. In the model s24, $r_{\rm sh}$ bounces to $\sim650$~km and then falls back, followed by a sequence of bounces and fallbacks. We will present more details of these two subclasses of models in Section~\ref{ssec:neu}. Afterwards, $\rho_c$ increases monotonically again, indicating a second steady accretion episode. In about $\sim0.5-1.5$ s, $\rho_c$ experiences a third plunge phase and the PCS finally collapses to a BH. The evolution of the PCS structure for the model s17 is shown in the middle panel of Figure~\ref{fig:profiles}. The profile with $\rho_{c14}=5.0$ is shortly before $t_{2c}$ and others are taken after the oscillation of $\rho_c$.  $Y_{q}$ also peaks off-center while the maximum value is $\sim0.75$ near BH formation. The smaller $Y_q$ is due to the lower temperature ($\sim50$~MeV) comparing to that in the model s33 ($\sim60$~MeV).
	
	For the two least compact models (s15 and s18), the changes in $\rho_c$, $r_{\rm sh}$ and $Y_{q,c}$ are less dramatic. For the model s18 with $\xi_{2.2}=0.20$, $\rho_c$ increases with a larger rate after $t_{2c}$, accompanied with shrinking of $r_{\rm sh}$ from 45~km to 35~km in $\sim0.2$~s. $Y_{q,c}$ increases from $\sim0.12$ to $\sim0.2$ in this episode. Then, the evolution of $\rho_c$, $r_{\rm sh}$ and $Y_{q,c}$ looks similar to that before $t_{2c}$. The sudden change of $Y_{q,c}$ at $\sim t_{1b}+1.9$~s is due to the coarse sampling of neutrino distribution function at large energies, which leads to problems in the transport when the neutrino chemical potential is high. The evolution of the PCS structure is shown in the right panel of Figure~\ref{fig:profiles}. It is similar to the other two models, but the maximum $Y_q$ is only $\sim0.5$ at the end of the simulation. The postbounce evolution of the model s15 with $\xi_{2.2}=0.13$ is steady and similar to that of CCSNe without the PT \citep[e.g. ][]{2018JPhG...45j4001O}. Until the end of our simulation ($t_{1b}+2.5$~s), these two models had not yet collapsed into BHs.

    In Table~\ref{tab:dyn}, we summarize the results of the postbounce dynamics for all simulated 21 models, listed with ascending $\xi_{2.2}$. Approximately, the effect of the PT on the PCS dynamics, i.e. the above 3 types of outcomes, is controlled by $\xi_{2.2}$. Consistent with the results in \cite{2011ApJ...730...70O,2020ApJ...894....4S}, the time interval between collapses is shorter for a larger $\xi_{2.2}$. We will relate the $\xi_{2.2}$ dependence to the properties of the hybrid EoS in Section~\ref{ssec:discuss}.

	\subsection{Neutrino signals \label{ssec:neu}}
	For failing CCSNe that lead to BH formation without a shock revival, their optical signals are expected to be faint or even absent \citep{1980Ap&SS..69..115N, Lovegrove_2013, 10.1093/mnras/sty306}. On the other hand, neutrinos and gravitational waves can be as visible as normal CCSNe \citep{2006PhRvL..97i1101S,2011PhRvL.106p1103O}. Our spherically-symmetrical simulations can predict the neutrino signals in these failing CCSN models with a hadron-quark PT.
	
	\begin{figure}[t!]
		\centering \includegraphics[width=0.47\textwidth]{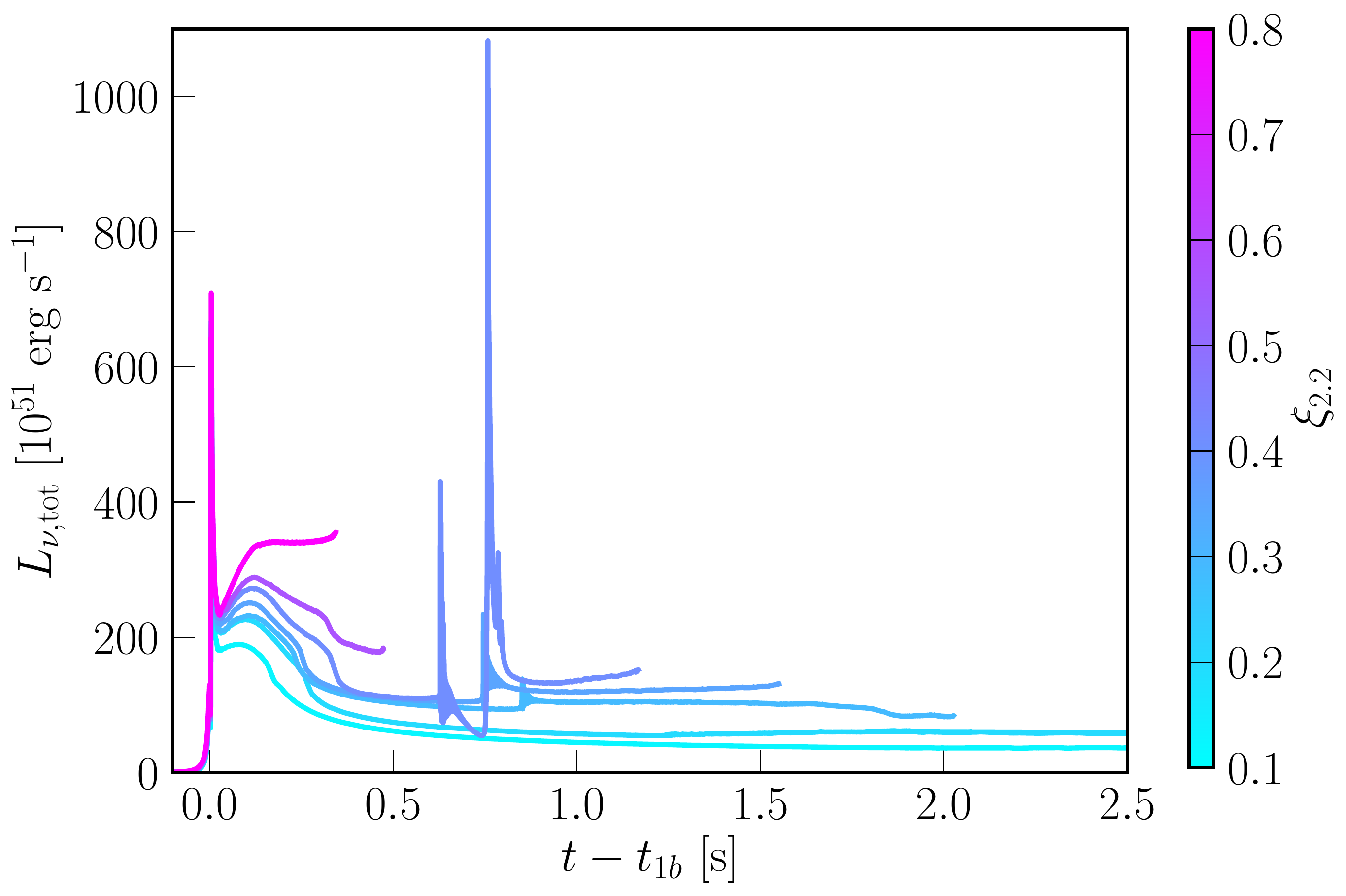}
		\caption{Same as Figure~\ref{fig:cen} but for the total neutrino luminosity vs. time after the first bounce. \label{fig:lnu}}
	\end{figure}

	 In Figure~\ref{fig:lnu}, we plot the total neutrino luminosity as a function of time after the first bounce for the same models as those in Figure~\ref{fig:cen}. Their evolution is closely related to the dynamics of the PCSs. For models with a large $\xi_{2.2}$ (s33 and s26), termination of neutrino signals are found at the time of BH formation. For models with a small $\xi_{2.2}$ (s15 and s18), the neutrino signals are similar to those in ordinary CCSNe \citep{2013ApJ...762..126O} because the PT does not induce any dramatic contraction of the PCSs. For models with an intermediate $\xi_{2.2}$ (s16, s17 and s24), the neutrino signals are oscillatory with a period of $\sim$ms after the second collapse, corresponding to the collapse, bounce, and oscillations of the PCSs. The oscillations last for $\sim50$ ms, in accord to the time duration of oscillations of $\rho_c$. 
	 
	 For the more compact progenitors in this intermediate set (e.g., s24 in Figure~\ref{fig:lnu}), there is a secondary burst in the neutrino luminosity after the second collapse (at $\sim0.13$~s for the s24 model). This burst has a larger luminosity than the neutronization burst after the first bounce. Its appearance follows the shock fallback and rebound in the middle panel of Figure~\ref{fig:cen}. In the following, we take a more detailed look into these models that experience a second bounce and oscillations.

	\begin{figure}[t!]
		\centering \includegraphics[width=0.47\textwidth]{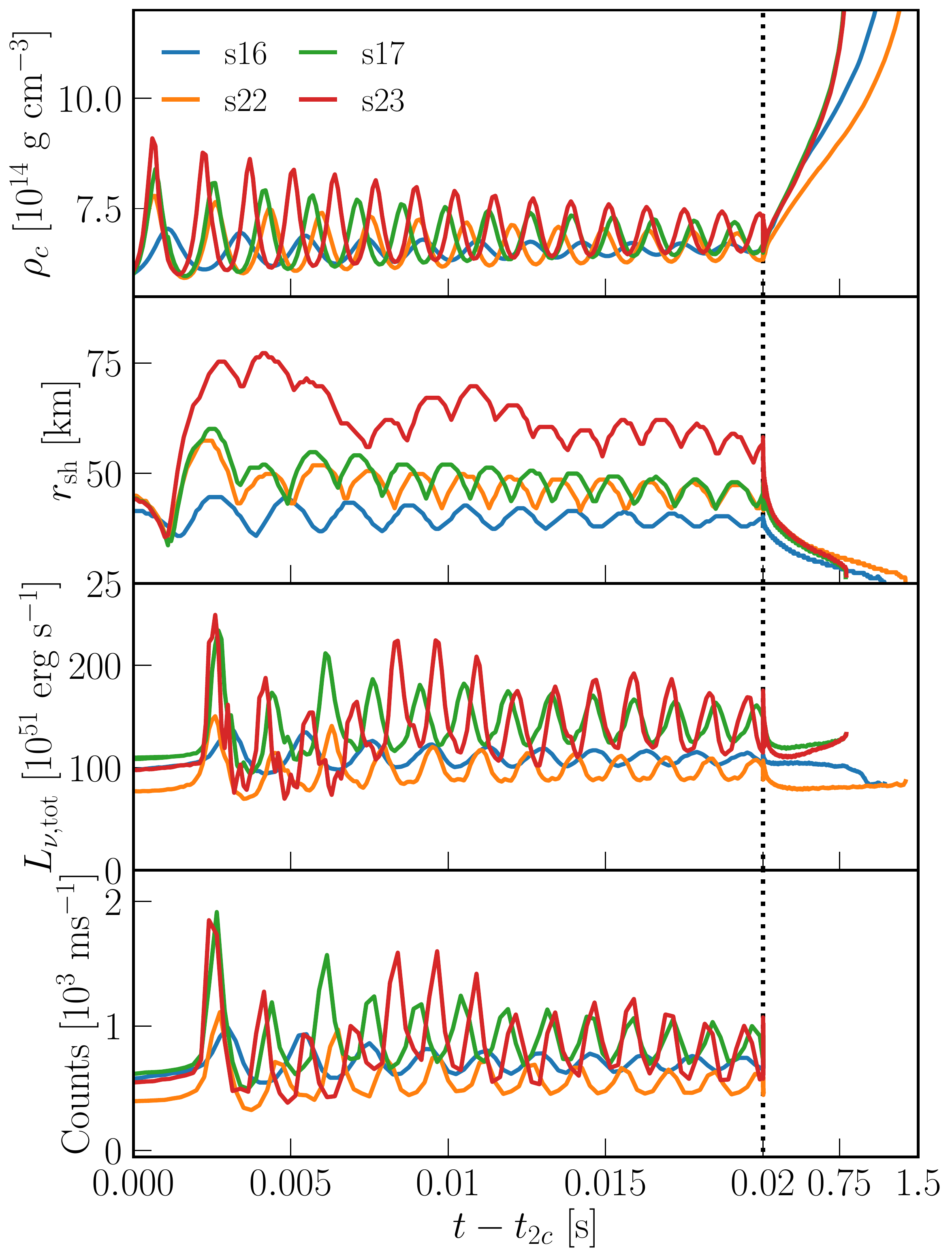}
		\caption{From top to bottom panel: central density, shock radius, total neutrino luminosity and neutrino event rate vs. time after the second collapse for 4 models with relatively small compactness. Note that timescale is different before and after $t-t_{2c}=0.02~\rm s$ (the black dotted line). Neutrino event rate is estimated for the IceCube detector assuming a source distance of 10~kpc and normal mass ordering for neutrinos.  \label{fig:class1}}
	\end{figure}

	In Figure~\ref{fig:class1}, we plot $\rho_c$, $r_{\rm sh}$, $L_{\nu, \rm tot}$ and neutrino event rate as a function of time after the second collapse for four models (s16, s22, s17, s23) with a relatively small $\xi_{2.2}$. In these simulations, $r_{\rm sh}$ only bounces up for several 10 km and then oscillates back and forth, with the same period as $\rho_c$ and $L_{\nu, \rm tot}$. The power of the bounce roughly increases with larger $\xi_{2.2}$. The oscillations last for $\sim50$~ms followed by a steady accretion phase. BH formation takes place in $\sim0.75-1.5$~s later, signaled by the quick increase of central density $\rho_c$, recession of shock radius $r_{\rm sh}$ and shut-off of neutrino emission $L_{\nu, \rm tot}$.
	
	To estimate the detectability of the oscillatory neutrino signals, we calculate the neutrino event rate for the IceCube \citep{Aartsen:2016nxy} detector (bottom panel of Figure~\ref{fig:class1}) using the SNOwGLoBES package \citep{doi:10.1146/annurev-nucl-102711-095006,Malmenbeck:2019Vr}. The calculation assumes a source distance of 10~kpc and normal mass ordering for neutrinos. \cite{2019PhRvD.100l3009W} derived a theoretical maximum distance for a $3\sigma$ detection of such a periodic neutrino signal by the IceCube detector (Eq.~5.2 in \cite{2019PhRvD.100l3009W}):
	\begin{equation}
	    d \sim 22.6{\rm kpc} \Big[\frac{\epsilon}{1}\Big] \Big[\frac{a}{0.25}\Big] \Big[\frac{A^{\rm 10kpc}}{400 {\rm ms}^{-1}}\Big]^{1/2} \Big[\frac{\Delta \tau}{10{\rm ms}}\Big]^{1/2},
	\end{equation}
	where $\epsilon$ is the purity of the signal, $a$ is the fractional amplitude of the periodic signal, $A^{\rm 10kpc}$ is mean neutrino event rate with a source distance of 10~kpc, and $\Delta\tau$ is the time interval of the periodic signal. Taking conservative values of $A^{\rm 10kpc}=800~{\rm ms}^{-1}$, $a=0.25$ and $\Delta\tau=10~{\rm ms}$ from the bottom panel of Figure~\ref{fig:class1} and assuming $\epsilon=0.5$, we get a detection distance of $\sim16$~kpc. If the inverted mass ordering is used, $A^{\rm 10kpc}$ is $\sim 400~{\rm ms}^{-1}$ and the detection distance is reduced to $\sim 11$~kpc. Therefore, we expect such a periodic neutrino signal is detectable in IceCube for a galactic CCSN.
	
	\begin{figure}[t!]
		\centering \includegraphics[width=0.47\textwidth]{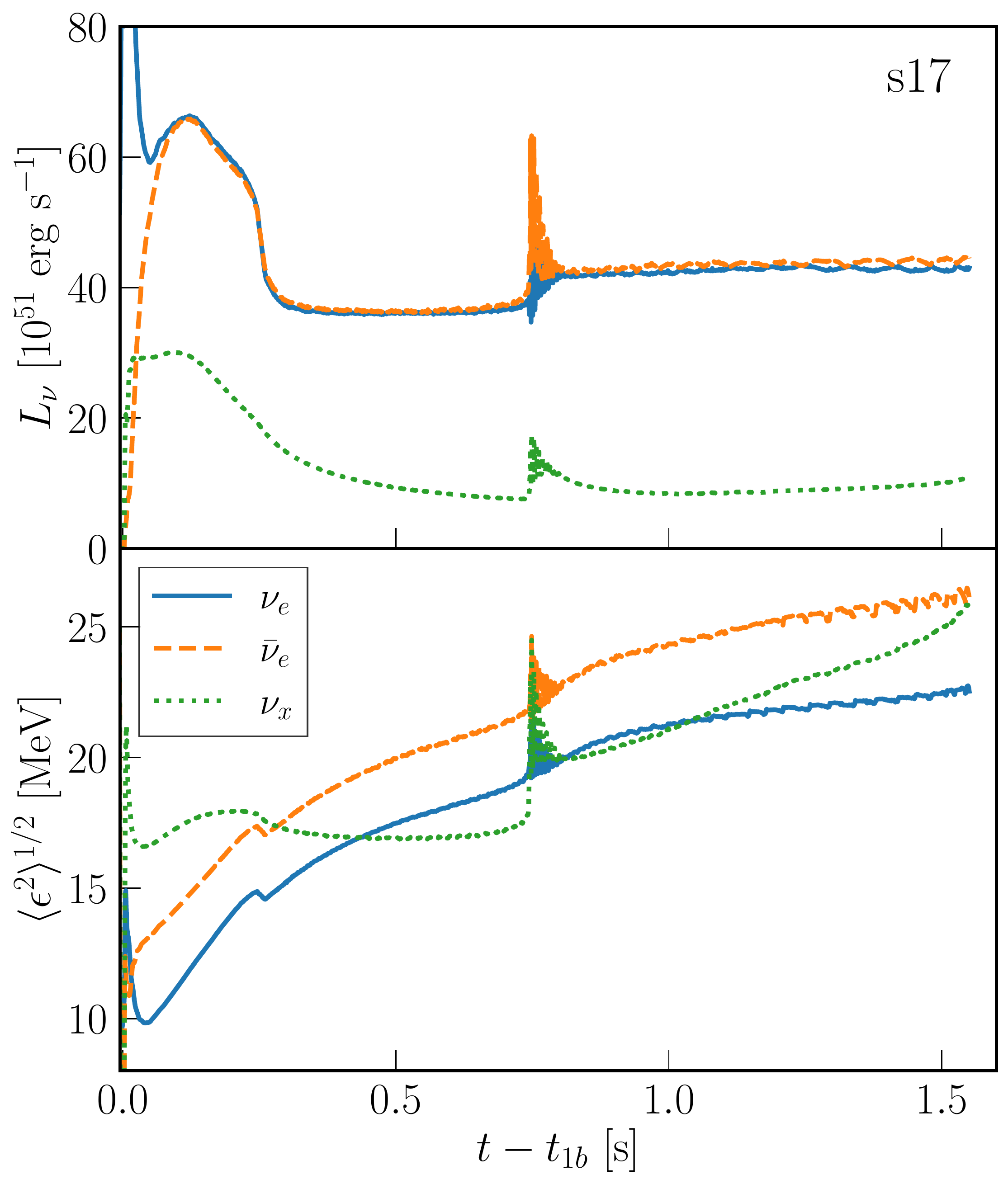}
		\caption{Neutrino luminosity (upper panel) and root-mean-squared energy (lower panel) for 3 flavors vs. time after the first bounce in model s17. \label{fig:s17neu}}
	\end{figure}
	
	We plot the flavor-dependent neutrino luminosity and root-mean-squared energy for the model s17 in Figure~\ref{fig:s17neu}. Before the second collapse, these signals look similar to those with pure hadronic EoSs. The drop of luminosity at $\sim t_{1b}+0.25$~s is due to accretion of the Si/Si-O shell interface and the sharp drop in the accretion rate on to the PCS. The second collapse and bounce of the PCS result in the increase of neutrino luminosity and energy for all three flavors, followed by oscillations with a period of $\sim$ms for $\sim50$~ms. Because of the earlier neutronization after the first bounce, electron anti-neutrinos ($\bar{\nu}_e$) have a larger luminosity than electron neutrinos ($\nu_e$) during this episode of oscillation.

	\begin{figure}[t!]
		\centering \includegraphics[width=0.47\textwidth]{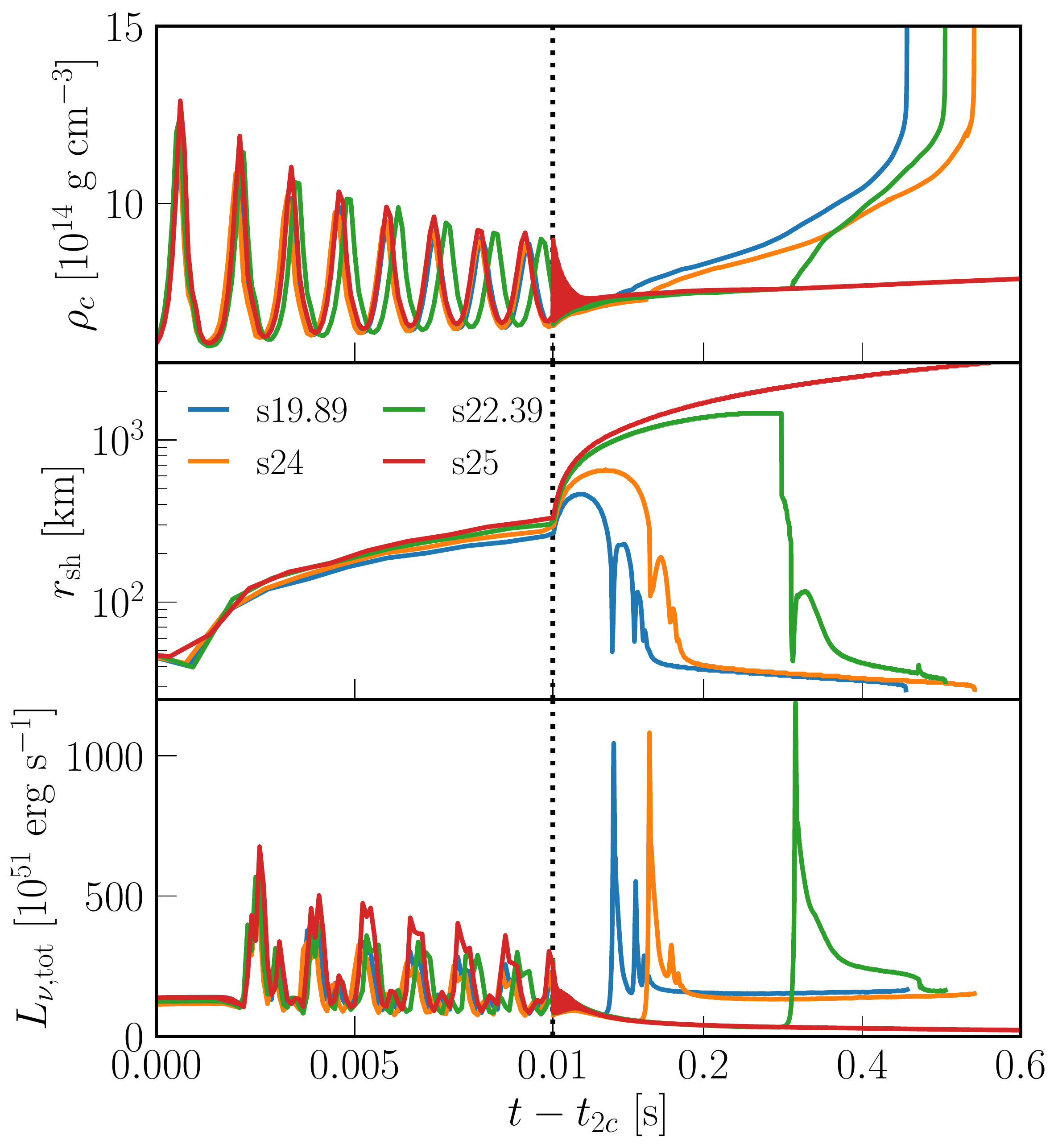}
		\caption{From top to bottom panel: central density, shock radius, and total neutrino luminosity vs. time after the second collapse for 4 models with relatively large compactness. Note that timescale is different before and after $t-t_{2c}=0.01~\rm s$ (the black dotted line). \label{fig:class2}}
	\end{figure}

	For models with a larger compactness, the bounce is more powerful and the accretion shock can expand to $>100$~km. In Figure~\ref{fig:class2}, we plot the postbounce evolution of $\rho_c$, $r_{\rm sh}$ and $L_{\nu, \rm tot}$ for 4 models (s19.89, s24, s22.39, s25) in this subclass. While $\rho_c$ and $L_{\nu, \rm tot}$ have similar oscillations as in the less compact models, $r_{\rm sh}$ expands to 100~km in $\sim2$~ms after the second bounce. For s19.89, s24 and s22.39, the shock reaches a maximum radius of $450$~km, $650$~km and $1450$~km, respectively. Then the shock experiences several cycles of fallback and bounce until a new steady accretion phase sets in. The shock bounce results in bursts of neutrinos with a total luminosity ($L_{\nu, \rm tot}$) comparable to, or even larger than, the neutronization burst shortly after the first bounce. \added{For the model s22.39, there is a sudden drop of $r_{\rm sh}$ from $\sim$1450~km to $\sim$450~km. This is because the original shock has a smaller velocity gradient than $\Delta v$ (criterion for finding a shock) and $r_{\rm sh}$ is switched to the position of an inner shock.} For the s25 model, the shock had not yet started to recede by the end of the simulation. However, there is no unbound matter with a positive total energy associated with the shock. Thus, we expect that the shock will still fall back onto the PCS for our spherically-symmetric simulation. The dynamics after the second bounce in models s22.31 and s22.38 are similar to s25 and the simulations terminate when $Y_e$ at $\sim20$~km exceeds the maximum $Y_e$ in the EoS table (0.6). Of all the models explored here, these are the closest to achieving an explosion. We also expect these models to be impacted by multidimensional effects.
	
	\begin{figure}[t!]
		\centering \includegraphics[width=0.47\textwidth]{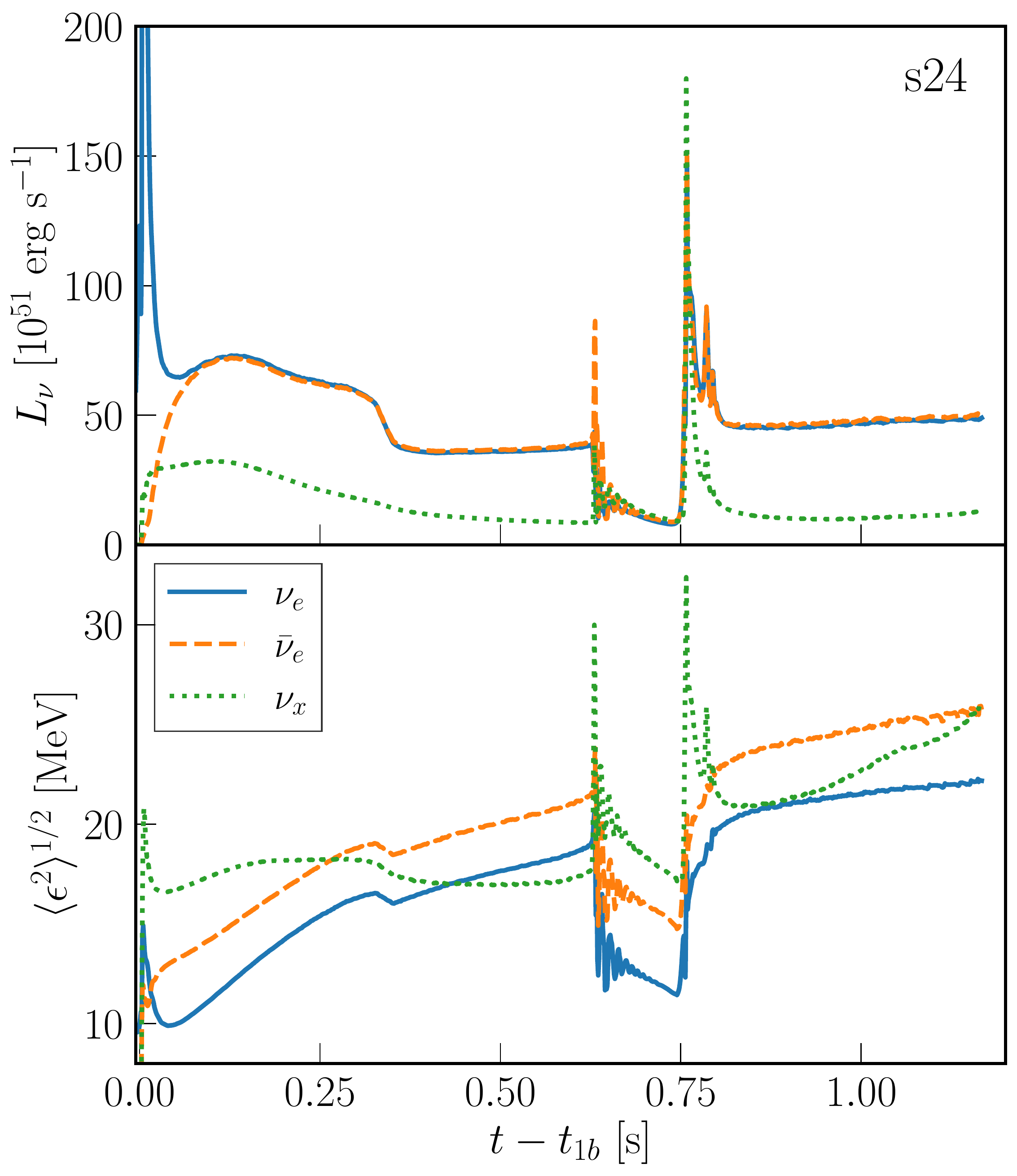}
		\caption{Same as Figure~\ref{fig:s17neu}, but for the model s24. \label{fig:s24neu}}
	\end{figure}
	
	We plot the flavor-dependent neutrino luminosity and root-mean-squared energy for the model s24 in Figure~\ref{fig:s24neu}. Again, they resemble those obtained with pure hadronic EoSs before the second collapse. The drop in luminosity at $\sim t_{1b}+0.3$~s corresponds to accretion of the Si/Si-O shell interface and the sharp drop in the accretion rate. Except for the first burst, the luminosity of electron-type neutrinos in the oscillatory episode is smaller than that before second collapse by a factor of 2.5 due to the shock expansion and drop of the accretion rate. In these spherically-symmetric simulations, whenever the accretion shock falls back and bounces, both neutrino luminosity and energy undergo related bursts. As the new accretion phase sets in, they return to the same level as that before the second collapse. The same behavior is also observed for models s19.89 and s22.39.
	
	In summary, the neutrino signal and its relation to the behavior of shock evolution can be a good indicator for the PT inside the PCS of failing CCSNe. In multidimensional simulations, coupling of the oscillations and convective motions inside the PCS may lead to GW emission with similar frequencies as found here for neutrino signals. It is an interesting extension to study the correlation between the neutrino and GW signals for the PT studied in our work.
	
    \subsection{Interpretation of the progenitor dependence \label{ssec:discuss}}

	\begin{figure}[t!]
		\centering \includegraphics[width=0.47\textwidth]{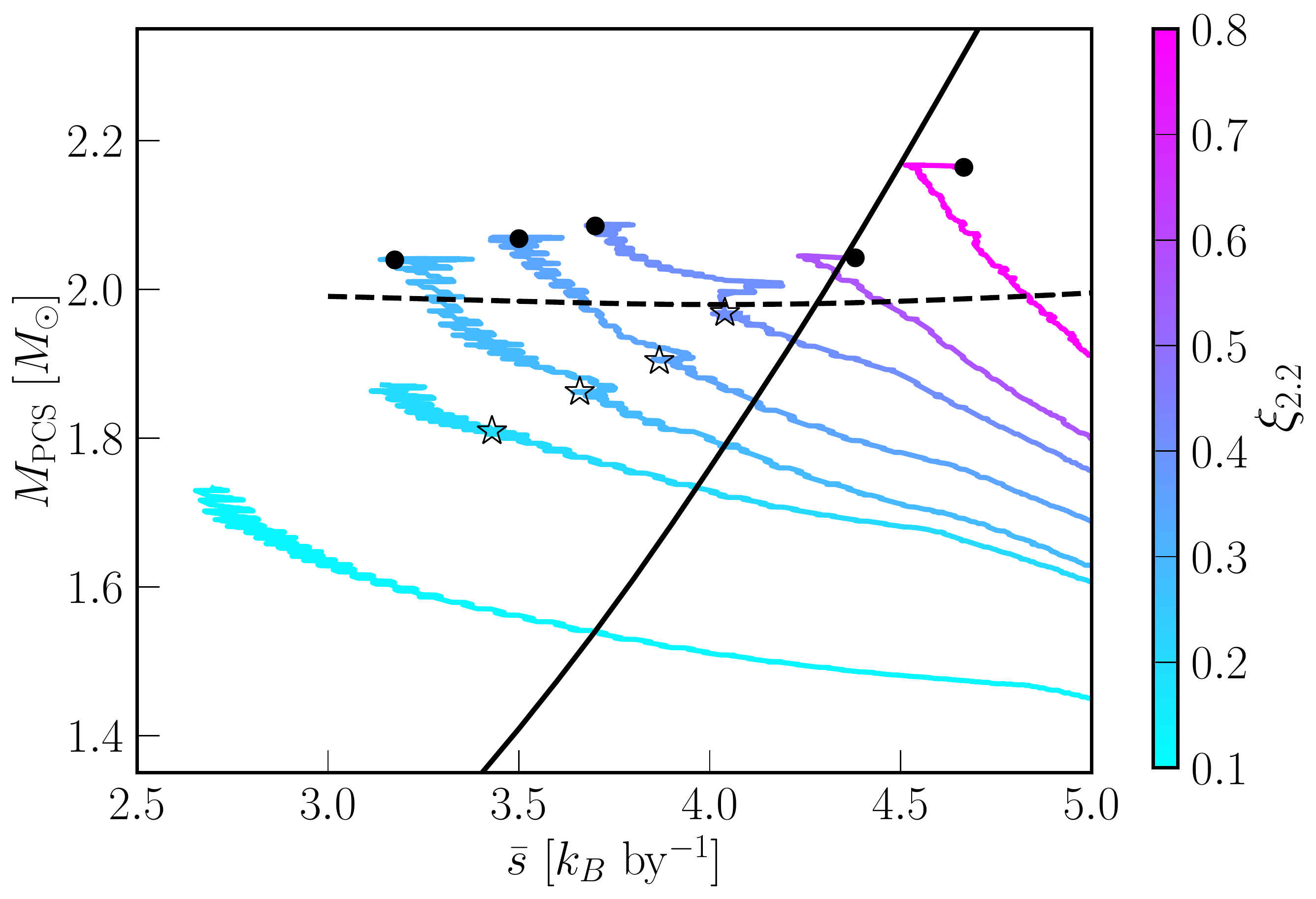}
		\caption{Evolution of the common entropy defined in Eq.~\ref{eq:sbar} and gravitational mass of the PCSs in models s15, s18, s16, s17, s24, s26, s33, from low to high compactness. The black unfilled stars and filled circles indicate the moments of the second collapse and BH formation, respectively. Black solid and dashes lines are the entropy dependent $M_{2,\max}$ and $M_{3,\max}$ in Figure~{\ref{fig:mr}}. \label{fig:s_mpns}}
	\end{figure}

	To relate the progenitor-dependent postbounce dynamics with the properties of the hybrid EoS, we analyse the PCS evolution in the mass-specific entropy diagram as \citet{2020ApJ...894....4S}. We compute the most common entropy of the PCS as
	\begin{equation}
	\bar{s} = \frac{\int_{0}^{M_0} s H(s(M)-s_{\min})  dM}{\int_{0}^{M_0}  H(s(M)-s_{\min})  dM}.   \label{eq:sbar}
	\end{equation}
	Here, $s(M)$ is the specific entropy per baryon at mass coordinate $M$. $s_{\min}$ is the minimum $s$ between the hottest part of the PCS and the accretion shock, lying at mass coordinate $M_0$. $H$ is the Heaviside step function
	\begin{equation}
	H(x) = \begin{cases}
	1 \quad {\rm if}~x\ge 0, \\
	0 \quad {\rm if}~x< 0. 
	\end{cases}
	\end{equation}
	This definition of $\bar{s}$ avoids the central cold region of the PCS as well the hot region right below the accretion shock. We opt to define the most common entropy somewhat differently than \cite{2020ApJ...894....4S} to avoid large entropy variations in the paths traced by each model in the $M_{\rm{PCS}}-\bar{s}$ diagram in our GR1D simulations.
	
	The evolution of the PCSs in the $M_{\rm PCS}$ and $\bar{s}$ diagram is shown in Figure~\ref{fig:s_mpns} for the same models as Figure~\ref{fig:cen}. By comparing the evolution of PCSs to the entropy dependent $M_{2,\max}$ and $M_{3,\max}$, the trend of progenitor dependence can be understood as follows. Progenitors with different $\xi_{2.2}$ produce hot PCSs with a different $\bar{s}$, whose M-R curves are similar to those in Figure~\ref{fig:mr}. Depending on the appearance of the third-family topology and the relative difference in $M_{2,\max}$ and $M_{3,\max}$, the outcome is different for a different $\bar{s}$. For a small $\xi_{2.2}$ and, thus, a small $\bar{s}$ without a $M_{2,\max}$, the PCS will collapse to a BH roughly once its mass exceeds $M_{3,\max}$. For a large $\xi_{2.2}$ and $\bar{s}$ where $M_{2,\max}\geq M_{3,\max}$, the PCS will collapse to a BH when its mass approximately exceeds $M_{2,\max}$.
	
	Particularly interesting models are those that lie in-between, i.e. $M_{2,\max}< M_{3,\max}$, the second collapse does not lead to BH formation immediately but is followed by a second bounce when the PCS mass exceeds $M_{2,\max}$. However, the bounce is not powerful enough to overcome the ram pressure of the materials outside the shock and to unbind the stellar envelope as in previous studies \citep{2009PhRvL.102h1101S,2018NatAs...2..980F,2020PhRvL.125e1102Z}. Instead, the PCS oscillates for tens of ms around the new equilibrium structure due to the excess kinetic energy. Afterwards, a new steady accretion phase sets in and the PCS will finally collapse into a BH when its mass overcomes $M_{3,\max}$.
	
	\begin{figure}
		\centering\includegraphics[width=0.47\textwidth]{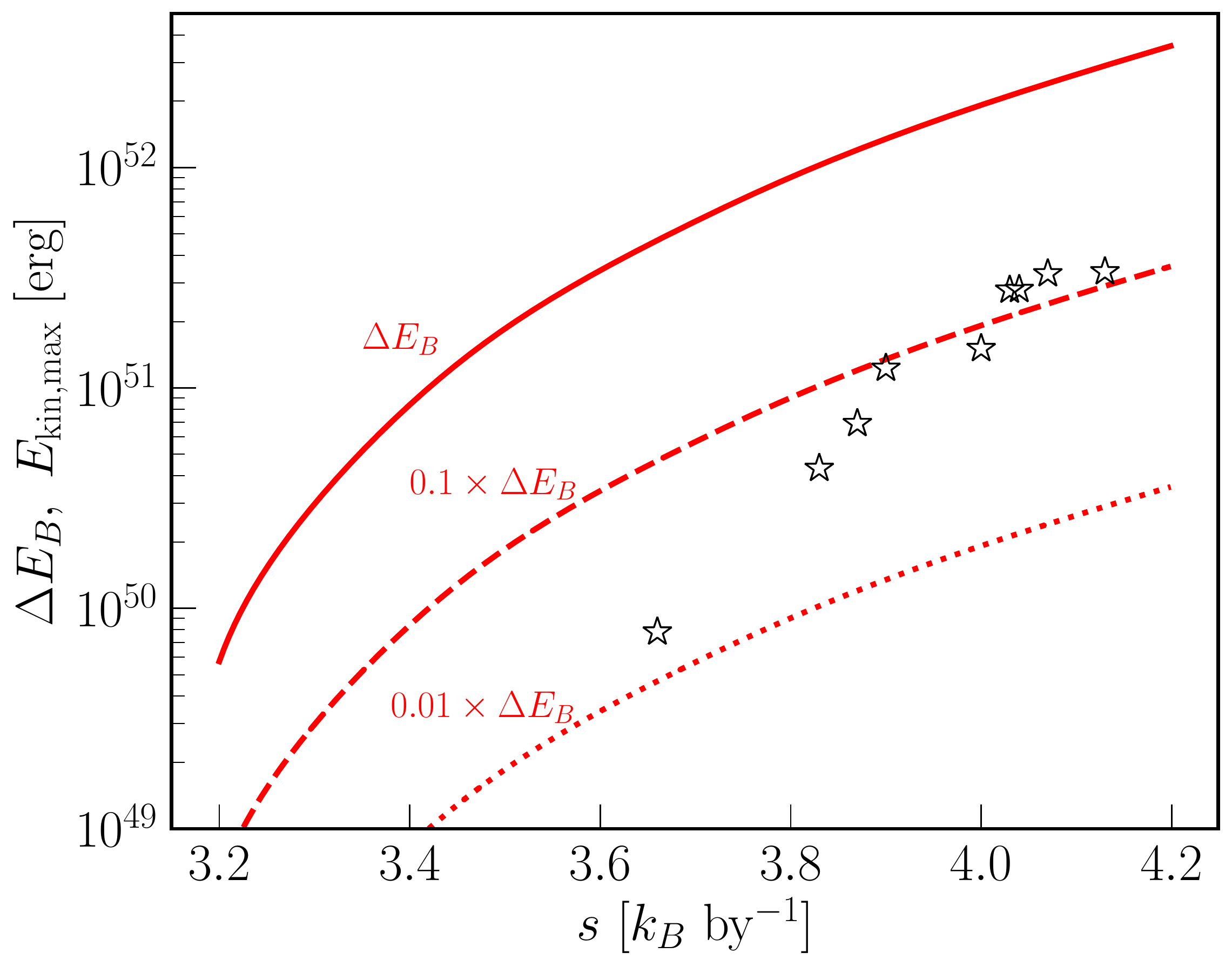}
		\caption{Difference in binding energies $\Delta E_{B}$ (Eq.~\ref{eq:difB}) as a function of  the specific entropy. Black unfilled stars are the maximum positive kinetic energies of the PCSs after the second bounce in models that the second collapse does not lead to BH formation. \label{fig:difB}}
	\end{figure}
	
	To have a quantitative view of the two different subclasses of shock expansion described in Section~\ref{ssec:neu} and to understand why the second bounce does not lead to shock revival, we estimate the binding energy released during the second collapse by a quantity $\Delta E_{B}$
	\begin{equation}
	\Delta E_{B} =  M_{2,\max} - M_{2,\max}^{'}. \label{eq:difB}
	\end{equation}
	Here, $M_{2,\max}^{'}$ is the gravitational mass of the compact star which has the same baryonic mass as $M_{2,\max}$ but is on the inner stable branch of the M-R curve (c.f. Figure~\ref{fig:mr}). We plot $\Delta E_{B}$ as a function of the specific entropy $s$ in Figure~\ref{fig:difB}, in comparison with the maximum positive kinetic energies $E_{\rm kin, \max}$ of the PCSs during the second bounce. Only part of the binding energy $\Delta E_{B}$ ($1-10\%$) transforms into kinetic energy $E_{\rm kin, \max}$, both of which follow an increasing trend with $s$. This explains why for larger $\xi_{2.2}$ the second bounce is more powerful and pushes up the accretion shock to a larger radius.
	
	For the hybrid EoS (\texttt{B165}) used in \cite{2009PhRvL.102h1101S} and \cite{2020PhRvL.125e1102Z} with $B=165$~MeV and $\alpha_s=0$, $M_{2,\max}$ emerges at a smaller entropy ($\sim1.5~k_B~{\rm by}^{-1}$) and $M_{2,\max}$ is larger than $M_{3,\max}$ for $s\gtrsim4~k_B~{\rm by}^{-1}$. $\Delta E_{B}$ is $\sim2\times10^{52}$~erg and $\sim1\times10^{53}$~erg at $s=3.0~k_B~{\rm by}^{-1}$ and $s=4.0~k_B~{\rm by}^{-1}$, respectively. These are $\sim$1-2 orders of magnitude larger than $\Delta E_{B}$ of the hybrid EoS employed in this work. Therefore, the \texttt{B165} EoS favors shock revival by the second bounce for progenitor models with a small compactness, as found in \cite{2009PhRvL.102h1101S} and \cite{2020PhRvL.125e1102Z}. \added{The larger $\Delta E_{B}$ for the \texttt{B165} EoS is due to the smaller radius of its third-family compact star. For a cold and $\beta-$equilibrated compact star with a gravitational mass of $1.4~M_\odot$, the \texttt{B165} (\texttt{B145}) EoS yields a radius of 9.06~km (13.0~km). The small radius for the \texttt{B165} EoS is incompatible with the constraint of compact-star radius with the GW170817 observation \citep{2017ApJ...850L..34B,2020Sci...370.1450D}. } We remark that the nature of such a PT is still uncertain \citep{2017RvMP...89a5007O} and may be unveiled by
    future multi-messenger observation of CCSNe and binary neutron star mergers.
	
	\section{Conclusions \label{sec:con}}
	In this paper, we have systematically investigated the progenitor dependence of a hadron-quark PT in failing CCSNe which result in stellar-mass BH formation. The PT is included in a hybrid EoS whose maximum mass of cold compact stars is $2.0~M_\odot$. It leads to an early second dynamical collapse of the PCS, and the outcomes for different progenitors as a function of their bounce compactness parameter $\xi_{2.2}$ are summarized as follows.
	\begin{enumerate}
		\item For $\xi_{2.2}\gtrsim0.51$, this second collapse directly leads to BH formation. Similar to \cite{2013A&A...558A..50N}, the PT shortens the time between the first bounce and BH formation. Shut-off of neutrino signals is expected as the PCS collapses to a BH.
		\item For $0.38\lesssim\xi_{2.2}\lesssim0.51$, a second bounce pushes the accretion shock to a large radius $\sim100~$km. During this bounce phase, neutrino emission is fainter than that before second collapse due to the drop of accretion rate. Nonetheless, the second bounce shock is not powerful enough to unbind the stellar envelope in our simulations. The shock experiences cycles of fallback and rebound. Bursts of neutrinos are observed accompanying the shock recession and expansion. Then, after a new steady accretion phase, BH formation takes place with the shut-off of neutrino emission.
		\item For $0.24\lesssim\xi_{2.2}\lesssim0.38$, the second bounce is weaker and only pushes the accretion shock up for tens of km, followed by oscillations of the shock radius and PCS. The neutrino emission is oscillatory with a period of $\sim$ms during this phase. Then a new steady accretion phase sets in and leads to BH formation. 
		\item For $\xi_{2.2}\lesssim0.24$, there is no dynamical collapse of the PCS and the postbounce dynamics is similar to the case without the PT.
	\end{enumerate}
	
	Taking the $\xi_{2.2}$ distribution\footnote{Here we use the progenitor compactness $\xi_{2.2}|_{t=0}$ at the onset of collapse and assume a one-to-one mapping from $\xi_{2.2}|_{t=0}$ to bounce compactness $\xi_{2.2}|_{t=t_{1b}}$ as listed in Table~\ref{tab:dyn}. } of the model set sxx (with zero-age-main-sequence mass ranging from 12 to 60 $M_\odot$) and the Salpeter initial mass function (IMF) with slope -2.35 \citep{1955ApJ...121..161S}, we get the probabilities for stars in the above four $\xi_{2.2}$ ranges to be $\sim 37\%$, $20\%$, $9\%$ and $34\%$, respectively. Therefore, $\sim 29\%$ of stars in this model set would emit oscillatory neutrino signals after the PT-induced second collapse and bounce. The percentage will be reduced by around half if taking the lower limit in the IMF to be 8 $M_\odot$. Note that the fraction is quite uncertain, subjecting to the uncertainties in the progenitor evolution, the hybrid EoS, and the explosion dynamics. We note that the sxx model set from \cite{2018ApJ...860...93S}, due to the low mass loss rate, contains a large number of high compactness progenitors.
	
	The progenitor dependence is well explained by the entropy dependent M-R curves of compact stars for the hybrid EoS. As a third-family branch emerges and becomes more prominent at a higher entropy, the M-R curves have two extreme masses, $M_{2,\max}$ and $M_{3,\max}$, connected by an unstable and a stable branch. When the mass of a PCS exceeds $M_{2,\max}$, it enters the secondary unstable branch and experiences a dynamical collapse. If $M_{3,\max}\le M_{2,\max}$, this second collapse will directly lead to BH formation. Otherwise if $M_{3,\max}>M_{2,\max}$, the PCS bounces and oscillates around the new equilibrium configuration after this second collapse. Because a larger amount of binding energy is released during the collapse of a hotter PCS, the power of the bounce increases for models with a larger $\xi_{2.2}$.
	
	The current study has been performed using a single hybrid EoS, but we expect the findings which relate to the third-family topology \citep{2013PhRvD..88h3013A,2016PhRvD..94j3001H} will be qualitatively the same for other EoSs including a similar PT. If the release of binding energy is large enough, the bounce of the PCS may be able to revive the supernova shock even in spherical symmetry \citep{2009PhRvL.102h1101S,2018NatAs...2..980F}. The next step is to systematically study the entropy dependent M-R relations for other hybrid EoSs. 
	
	Another extension of this paper is to explore multidimensional effects, which can lead to GW emission through coupling of the PCS oscillations to its convective motions \citep{2020PhRvL.125e1102Z}. Models with $\xi_{2.2}$ in-between 0.45 and 0.51, for which the second bounce shock can expand to $\sim1000$~km, may successfully explode in multidimensional simulations, as in the case of non-hybrid EoSs. Also, we expect the periodic neutrino signals will be modulated by multidimensional effects such as rotation \citep{2019PhRvD.100l3009W} and hydrodynamical instabilities \citep{2019ARNPS..69..253M}.
	
	\acknowledgments 
    We thank Irina Sagert for making the hybrid equation of state publicly available on the CompOSE website. This work is supported by the Swedish Research Council (Project No. 2018-04575 and 2020-00452). The simulations were enabled by resources provided by the Swedish National Infrastructure for Computing (SNIC) at PDC and NSC partially funded by the Swedish Research Council through grant agreement No. 2016-07213. SZ acknowledges the kind hospitality of Tsung-Dao Lee Institute during the preparation of this manuscript.
    
    \software{GR1D
    \citep{2010CQGra..27k4103O,2015ApJS..219...24O}, Numpy \citep{harris2020array}, Matplotlib \citep{Hunter:2007} }

     \newpage
    
	\bibliographystyle{aasjournal}
	\bibliography{the_bib}

\end{document}